
----------------------------------------------------------------------------
\overfullrule=0pt
\magnification 1200
\baselineskip=18 true bp
\def\jpa #1 #2 #3 {{\sl J. Phys.\ A} {\bf #1}, #2 (#3)}
\def\pra #1 #2 #3 {{\sl Phys.\ Rev.\ A} {\bf #1}, #2 (#3)}
\def\pre #1 #2 #3 {{\sl Phys.\ Rev.\ E} {\bf #1}, #2 (#3)}
\def\prl #1 #2 #3 {{\sl Phys.\ Rev.\ Lett.} {\bf #1}, #2 (#3)}
\def\pA #1 #2 #3 {{\sl Physica A} {\bf #1}, #2 (#3)}
\def\zw #1 #2 #3 {{\sl Z. Wahrsch.\ verw.\ Gebiete} {\bf #1}, #2 (#3)}
\def\jsp #1 #2 #3 {{\sl J. Stat.\ Phys.} {\bf #1}, #2 (#3)}
\newcount\eqnum \eqnum=0  
\def\eqnoi{\global\advance\eqnum by 1\eqno(\the\eqnum)}
\newcount\refnum\refnum=0  
\def\refi{\smallskip\global\advance\refnum by 1\item{\the\refnum.}}
\def\pd#1#2{{\partial #1\over\partial #2}}      
\def\td#1#2{{d #1\over d #2}}      
\def\Ct{C(t)}

\def\Cmt{C_m(t)}
\def\Cr{C(r)}
\def\Cmr{C_m(r)}
\def\Crt{C(r,t)}

\def\Fmr{F_m(\rho)}
\def\lap{D\Delta}

\def\dx{\Delta x}
\def\prt{P(r,t)}
\def\logt{\log(t)}
\def\logr{\log(r)}
\def\Ai{A_i}
\def\Aj{A_j}
\def\Ak{A_k}
\def\Aij{A_{i+j}}
\def\Aijk{A_{i+j+k}}
\def\Kij{K(i,j)}
\def\Kijk{K(i,j,k)}
\def\Nt{N(t)}
\def\Kn{K(i_1,\cdots,i_n)}
\def\g{\Gamma}
\def\a{\alpha}
\def\b{\beta}


\centerline{\bf Diffusion-Limited Aggregation Processes with 3-Particle
Elementary Reactions}
\bigskip
\centerline{\bf P.~L.~Krapivsky}
\smallskip
\centerline{Center for Polymer Studies and Department of Physics}
\centerline{Boston University, Boston, MA 02215}
\vskip 1in
\centerline{ABSTRACT}
{\narrower\narrower\smallskip\noindent
A diffusion-limited aggregation process, in which clusters coalesce
by means of 3-particle reaction, $A+A+A \to A$, is investigated.
In one dimension we give a heuristic argument that predicts
logarithmic corrections to the mean-field asymptotic behavior
for the concentration of clusters of mass $m$ at time $t$,
$\Cmt \sim m^{-1/2}(\logt/t)^{3/4}$, for $1 \ll m \ll \sqrt{t/\logt}$.
The total concentration of clusters, $\Ct$, decays as
$\Ct\sim \sqrt{\logt/t}$ at $t \to \infty$. We also investigate
the problem with a localized steady source of monomers and find
that the steady-state concentration $\Cr$ scales as
$r^{-1}(\logr)^{1/2}$, $r^{-1}$, and $r^{-1}(\logr)^{-1/2}$,
respectively, for the spatial dimension $d$ equal to 1, 2, and 3.
The total number of clusters, $\Nt$, grows with time as
$(\logt)^{3/2}$, $t^{1/2}$, and $t(\logt)^{-1/2}$ for $d$ = 1, 2,
and 3. Furthermore, in three dimensions we obtain an
asymptotic solution for the steady state cluster-mass
distribution: $\Cmr \sim r^{-1}(\logr)^{-1}\Phi(z)$,
with the scaling function $\Phi(z)=z^{-1/2}\exp(-z)$ and
the scaling variable $z \sim m/\sqrt{\logr}$.
}
{
\narrower\bigskip\noindent
P. A. C. S. Numbers: 82.20.-w, 05.40.+j., 02.50.-r, 82.70.-y.
}

\vfill\eject
\medskip\centerline{\bf I. Introduction}\smallskip

Diffusion-limited aggregation processes have attracted
considerable recent interest in many fields of science
and technology [1]. Typically, aggregation processes can
be described by the binary reaction scheme
$$
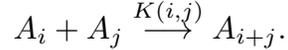
\Ai + \Aj \buildrel \Kij \over \longrightarrow \Aij.
$$
Here $\Ai$ denotes a cluster consisting of $i$ monomers, an $i$-mer,
and $\Kij$ is the rate at which the reaction between an $i$-mer
and a $j$-mer proceeds. Much of the understanding of the kinetics of
binary aggregation processes is based on the analysis of rate
equations and their exact and scaling solutions [1].
For sufficiently low dimensions, the diffusion mechanism
is not efficient enough and fluctuations in the densities
of diffusing reactants result in dimension-dependent
kinetic behavior at long times [2].

In a view of the richness of the kinetic behavior observed in the
bimolecular model, it is of interest to investigate more
complicated many-particle diffusion-limited aggregation processes.
In the present study, we focus on the 3-particle reaction scheme,
$$
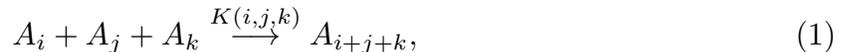
\Ai + \Aj + \Ak \buildrel \Kijk \over \longrightarrow \Aijk, \eqnoi
$$
both for homogeneous and inhomogeneous situations,
and outline a generalization for the $n$-particle case.

We will study the simplest $n$-particle aggregation process for
which both the reaction rates and diffusion coefficients do not
depend on masses of clusters, $\Kn$ = const and
$D_k$ = const. A notable feature of this process is that it
reduces to a simple chemical reaction scheme, $nA \rightarrow A$,
if one considers only the concentration of clusters.
Hence in the rate equation description, the concentration,
$\Ct$, obeys
$$
\td C t=-\gamma C^n, \eqnoi
$$
with $\gamma$ being the rate constant. In the long-time limit,
the concentration behaves as
$$
\Ct \simeq \bigl[\gamma (n-1)t\bigr]^{-1/(n-1)}. \eqnoi
$$

However, the mean-field rate equation approach
provides an accurate description of the kinetics
only above the upper critical dimension, $d_c$;
when $d \le d_c$, the mean-field theory does not predict
correct long-time behavior. For the aggregation model
with constant reactivities and diffusivities
the upper critical dimension is known: $d_c = 2/(n-1)$ [3-5].
Thus, for the binary reaction in one dimension
the kinetics is anomalous and the concentration decays as
$t^{-1/2}$ (see, e.g., [2] and references therein)
while the mean-field result is $\Ct \sim t^{-1}$.
On the other hand, for $n \ge 4$ the mean-field answer (3)
gives a correct asymptotic description of the kinetics
in one dimension. The 3-particle case is marginal
in one dimension and hence a logarithmic correction
has been expected [3-5].
After a number of attempts [3-6], the logarithmic
correction of the form $\Ct\sim\sqrt{\logt/t}$
has indeed been observed in a very recent study [7].

In the next Section II, we explore the diffusion-limited
3-particle aggregation process (1) in one dimension.
We justify heuristically the appearance of the logarithmic
correction for the concentration of clusters $\Ct$.
Moreover, we obtain a complete scaling description of the
cluster-mass distribution function: $\Cmt \sim (\logt/t)\Phi(z)$,
with the scaling function $\Phi(z)=z^{-1/2}\exp(-z)$ and
the scaling variable $z \sim m\sqrt{\logt/t}$.

In the last Section III, we examine the 3-particle
aggregation process with a spatially localized source
of monomers. We show that the system reaches the
steady state for arbitrary spatial dimension $d$.
We present evidence of two critical dimensions,
$d_c = 1$ as in the homogeneous system and $d^c = 3$
which demarcates the pure diffusive regime ($d>3$)
from the diffusion-reaction regime ($d \le 3$).
In three dimensions, we derive a complete asymptotic
solution for the steady state cluster-mass distribution.

\bigskip\centerline{\bf II. Aggregation process on a line}\smallskip

Consider a linear lattice on which point clusters undergo a
random walk. If $n$ clusters happen to occupy the same lattice
point, they aggregate irreversibly into a single cluster whose
mass is equal to the sum of the masses of $n$ parent clusters.
This is the $n$-body particle coalescence model (PCM) [8].
The binary PCM in one dimension has been solved by Spouge [9],
see also [10-14] for other exact solutions of several
generalization of the PCM. Note that we must keep
the lattice spacing $\dx$ finite, even in one dimension,
since otherwise the reaction will be absent in the $n$-particle
model for all $n \ge 3$. The possibility of passing to the
continuum limit in the binary PCM significantly simplifies
the analysis (see, e.g., Refs.~[10,11,14]), while for $n \ge 3$
the spacing $\dx$, which may be considered as the size of particles,
appears in the final results.

A simple heuristic argument explains the logarithmic corrections
for the 3-body PCM. Let $T$ be a typical time between successive
3-particle collisions in which cluster takes place. So the reaction
rate is proportional to $C/T$,
$$
\td C t \simeq - {C \over T}. \eqnoi
$$
We will estimate $T$ as follows: Consider a reference
frame at rest with respect to an arbitrary ``target'' particle.
When two other particles will be at the origin
simultaneously, the target particle will die.
Now consider all possible pairs of original particles.
In the following, we will refer to these pairs as
{\it imaginary} particles. For any pair, let us choose
the location of one partner as the $x$-coordinate
and the other location as the $y$-coordinate
of the corresponding imaginary particle.
Thus, we map the original diffusion process on the 1D
lattice onto a diffusion-like process on the 2D square
lattice with the same lattice spacing $\dx$.
Although imaginary particles do not undergo
a simple random walk, we shall assume that the asymptotic
behavior of this diffusion process is similar to the one
encountered in the simplest 2D random walk.

Now let us estimate the collision time $T$ by considering an
idealized 2D simple random walk of imaginary particles.
The target particle will die when some imaginary
particle will arrive at the origin.
The density $\prt$ of imaginary particles is governed by
$$
\pd{\prt}{t}=\lap \prt, \eqnoi
$$
with the initial condition
$$
P(r,t=0)=C^2 \eqnoi
$$
indicating the obvious fact that the density of imaginary particles
is just the square of the density of original ones, and with
the adsorbing boundary condition
$$
P(r=\dx,t)=0, \eqnoi
$$

A simple way to find an approximate solution of Eqs.~(5)-(7)
is to use a quasistatic approximation (see, e.g., [15]).
In this approximation one solves the steady diffusion
equation and accounts for the time dependence by a moving
boundary condition. This very simple approach often gives
asymptotically exact results (see, e.g., [16]). In the
present problem, the quasistatic approximation yields
$$
\prt=C^2 {\log(r/\dx)\over\log(\sqrt{Dt}/\dx)}.\eqnoi
$$

Now the collision time $T$ may be evaluated by computing
the flux to the origin and then by equating the flux to the unity:
$$
\int\limits_0^T 2\pi D \pd {P(r=\dx,t)}{r} dt = 1. \eqnoi
$$
This gives the final estimate,
$T\simeq\log(1/C\dx\sqrt{2\pi})/2\pi DC^2$.
Substituting this result into (4), we arrive at the closed-form
approximate differential equation for the concentration of clusters:
$$
\td C t \simeq{2\pi DC^3\over\log(C\dx\sqrt{2\pi})}.\eqnoi
$$
Solving this equation at the long time limit one gets
$$
\Ct \simeq \sqrt{\log(2Dt/\dx^2)\over 8\pi Dt}. \eqnoi
$$

This result confirms previous suggestions [3-6] of possible
logarithmic corrections to the mean-field power-law decay.
Moreover, it is in a complete agreement with recent
simulational results [7]. A similar logarithmic correction
has been found in the binary PCM in two dimensions,
$\Ct\sim {\log(Dt/\dx^2)/Dt}$ [17,8,18].
{}From these findings for 2- and 3-body PCM in their critical
dimensions, $d_c = 2$ and $d_c = 1$, respectively,
one can expect the appearance of similar
logarithmic corrections for the $n$-body
PCM at the corresponding critical dimension $d_c = 2/(n-1)$:
$\Ct\sim (\log(Dt/\dx^2)/Dt)^{1/(n-1)}$.
Very recently, a similar result has been established
by a renormalization group calculation for the $n$-particle
annihilation process, $nA \to 0$, see [19].
Notice also that for $d<d_c$ dimensional analysis
suggests a power-law decay with the exponent dependent
on $d$ and independent on $n$, $\Ct\sim (Dt)^{-d/2}$.

Let us now consider the 3-body aggregation process (1).
On the mean-field level, the process is described by the
(generalized) Smoluchowski equation. For constant reactivities,
the Smoluchowski equation has the form:
$$
\td {C_m} t= \sum_{i+j+k=m}C_iC_jC_k - 3C_mC^2, \eqnoi
$$
with $C=\sum\nolimits_{i=1}^{\infty}C_i$. Notice that in Eq.~(12)
we set the reaction rate equal to 1 by an appropriate choice of units.

For the monodisperse initial conditions, $C_i(0)=N_0\delta _{i 1}$,
one can find the solution to Eq.~(12), see [4].
The densities of even-mass clusters vanish due to our choice of
initial data, while the odd-mass densities are
$$
C_{1+2m}=C_1{\g(m+1/2)\over\g(1/2)\g(m+1)}
   \Bigl(1-{C\over N_0}\Bigr)^m,  \eqnoi
$$
where $\g$ is the gamma function. For example, the concentrations
of monomers and clusters are given by
$$
C_1=N_0(1+4N_0^2t)^{-3/4},\qquad C=N_0(1+4N_0^2t)^{-1/2}. \eqnoi
$$

Note that in the long-time limit the cluster-mass distribution
(13) exhibits a scaling behavior of the form
$$
C_{1+2m}\simeq {C^2\over N_0}{\exp(-z)\over\ \sqrt{\pi z}}, \eqnoi
$$
with the scaling variable $z$, $z=mC/N_0$. Equations (13) and (15)
show that the concentration decays as $C_{1+2m} \sim m^{-1/2}t^{-3/4}$
for $1 \ll m \ll \sqrt{t}$. The exponents describing the
scaling behavior might be obtained without appealing to
the complete solution. For example, the exponent describing
the time dependence, $C_{1+2m} \sim t^{-3/4}$, can be derived
by solving Eq.~(12) directly for small $m$.

In analogy with the analysis of the chemical reaction scheme,
$3A \rightarrow A$, at the upper critical dimension one can
suppose that the aggregation reaction process (1) at $d=d_c=1$
should be described by improved rate equations:
$$
L\td {C_m} t= \sum_{i+j+k=m}C_iC_jC_k - 3C_mC^2, \eqnoi
$$
with $L$ being the logarithmic factor,
$$
L={\log(1/C\dx \sqrt{2\pi})}/(\pi D).
$$
Upon summing equations (16) over all $m$,
one can reproduce Eq.~(10) for the cluster concentration,
thus providing a useful check of self-consistency.

By applying a generating function technique, one can find an
asymptotic solution to the modified Smoluchowski equation (16).
The concentrations are given formally by the same expression (13)
but with the modified monomer density
$$
C_1\simeq{N_0}^{-1/2}\biggl[{\log(2Dt/\dx^2)\over 8\pi Dt}\biggr]^{3/4}.
\eqnoi
$$
Both the scaling form and the scaling variable are identical
to the corresponding mean-field result (15),
although the normalization factors and widths
of distributions differ by logarithmic factors.

A straightforward generalization to the $n$-body PCM shows that
at the critical dimension, $d=d_c = 2/(n-1)$, the cluster-mass
distribution exhibits the scaling behavior of the mean-field form [4]:
$$
C_{1+(n-1)m}(t)\simeq {C^2\over N_0\g({1\over n-1})}z^{-{n-2\over n-1}}e^{-z},
\eqnoi
$$
with the scaling variable $z$,
$$
z=mC/N_0 \sim m(\log(Dt/\dx^2)/Dt)^{1/(n-1)}.
$$
In particular, the concentration of momomers decays as
$$
C_1(t)\sim \biggl[{\log(Dt/\dx^2)\over Dt}\biggr]^{n/(n-1)^2}. \eqnoi
$$

Thus for $n$-body PCM we have obtained the scaling description
of the cluster-mass distribution function. Our approach is
heavily based on the assumption of equal diffusivities of the
reactants. [The second assumption of equal reactivities
directly follows from the assumption of equal diffusivities
since ``sizes'' of clusters in the PCM are the same].
However, from our results one can forecast some
features of the general case. Consider, for example, the 3-body
PCM with the cluster diffusion coefficients $D_m$ related
to their mass $m$ by $D_m \sim m^{-\delta}$. Since the typical
mass of the cluster grows inversely proportional to the
total concentration of clusters, $m_{\rm typ} \sim C^{-1}$,
one can insert an estimate for the typical diffusion
coefficient, $D_{\rm typ} \sim m_{\rm typ}^{-\delta} \sim
C^{\delta}$, into expression (11) for the total
concentration of clusters. This yields the following heuristic
estimate for $\Ct$,
$$
\Ct \sim \biggl[{\log(t)\over t}\biggr]^{1/(2+\delta)}, \eqnoi
$$
but such an approach cannot predict the scaling form of
the cluster-mass distribution function.

\bigskip\centerline{\bf III. Aggregation process with a localized
steady source}\smallskip

In this section we consider $n$-body PCM with a steady spatially
localized monomer input. We again focus on the 3-particle case and
also write some final results for arbitrary $n$. Let the source
of monomers is placed at the origin of $d$-dimensional space
and $J$ is the strength of the source.
Within the continuum approximation, the concentration of clusters
at time $t$ and at distance $r$ from the source, $\Crt$,
satisfies the reaction-diffusion equation
$$
\pd{\Crt}{t}=\lap\Crt - DR + J\delta ({\bf r}), \eqnoi
$$
Here $DR$ is the reaction term:
$R=C^3, -C^3/\log(C), and C^{1+2/d}$ for $d>1, d=1$, and $d<1$,
respectively. For $d>1$, the reaction term has the mean-field form;
for the marginal case $d=1$, the form of the reaction term has
been derived in Section 2 (we ignore numerical factors and set
$\dx=1$ by an appropriate choice of the length scale);
and for $d<1$, one can choose the reaction term
of the form $C^{1+2/d}$ since it gives the correct long-time
decay for the homogeneous system, $\Ct \sim (Dt)^{-d/2}$.

For the binary PCM, a qualitative investigation of similar
reaction-diffusion equation has been performed in Ref.~[12].
Following the same line of reasoning, let us assume that the
system reaches the steady state and try a solution of the
power-law form. By inserting such a form to the governing
equation one finds that dimension $d=3$ also plays a role
of critical dimension in the present problem. For $d>3$,
clusters do not interact far away from the source and
the concentration decays as $r^{-(d-2)}$, i.e.,
as in the limit of no reaction.
The reaction leads to the renormalization of the strength
of the source but does not change the behavior qualitatively.
In other dimensions, the reaction is relevant at all scales,
and far away from the source $\Cr$ behaves as
$$
\Cr \sim \cases {r^{-1}(\logr)^{-1/2}, &if $d=3$;\cr
                 r^{-1}, &if $1<d<3$;\cr
                 r^{-1}(\logr)^{1/2}, &if  $d=1$;\cr
                 r^{-d}, &if $d<1$. }               \eqnoi
$$
The critical cases $d=3$ and $d=1$ have been treated separately.
Since we expected logarithmic corrections to the power-law behavior
$\Cr \sim r^{-1}$ for $1<d<3$, we tried a solution of the form
$\Cr \sim r^{-1}(\logr)^{\nu}$, inserted this form into Eq.~(21),
and finally found the exponent $\nu$ by asymptotically equating
the most significant terms.

Notice that the system with a localized source but without reaction
reaches the steady state only for sufficiently large spatial
dimension, $d>2$. On the other hand, the system with reaction
reaches the steady state for any $d$. Notice also the appearence of
logarithmic corrections to the power-law behavior in two critical
dimensions, $d_c=1$ and $d^c=3$: the former results from the
logarithmic factor in the reaction term, while the latter reflects
the fact that at $d=d^c$, the reaction just becomes relevant.

Since clusters perform a random walk and since the source was turned on
at $t=0$, clusters will propagate diffusively up to the distance
of the order $\sqrt{t}$. Therefore at $r<\sqrt{t}$ the concentration
of clusters approaches to the steady state limit given by Eq.~(22);
on the other hand, $\Cr \to 0$ rapidly for $r>\sqrt{t}$. Hence,
the total number of clusters, $\Nt$, may be estimated from the
relation $\Nt \sim \int_{0}^{\sqrt t}r^{d-1}\Cr dr$. This yields
$$
\Nt \sim \cases {t, &if $d>3$;\cr
                 t(\logt)^{-1/2}, &if $d=3$;\cr
                 t^{(d-1)/2}, &if $1<d<3$;\cr
                 (\logt)^{3/2}, &if $d=1$;\cr
                 \logt, &if $d<1$.\cr}               \eqnoi
$$

A straightforward generalization to the $n$-body PCM shows that,
again, two critical dimensions $d_c$ and $d^c$ demarcate
different behaviors: $d_c=2/(n-1)$ and $d^c=2n/(n-1)$.
For the steady-state cluster concentration one finds
$$
\Cr \sim \cases{r^{-(d-2)}, &if $d>d^c$;\cr
                r^{-d_c}(\logr)^{-d_c/2}, &if $d=d^c$; \cr
                r^{-d_c}, &if $d_c<d<d^c$;\cr
                r^{-d_c}(\logr)^{d_c/2}, &if $d=d_c$;\cr
                r^{-d}, &if $d<d_c$;\cr}               \eqnoi
$$
while the total number of clusters scales as
$$
\Nt \sim \cases {t, &if $d>d^c$;\cr
                 t(\logt)^{-d_c/2}, &if $d=d^c$;\cr
                 t^{(d-d_c)/2}, &if $d_c<d<d^c$;\cr
                 (\logt)^{d^c/2}, &if $d=d_c$;\cr
                 \logt, &if $d<d_c$.\cr}               \eqnoi
$$

Consider now the behavior of the steady state concentrations, $\Cmr$,
for the 3-body PCM in the most interesting three dimensional case.
The concentrations $\Cmr$ obey the reaction-diffusion equations
$$
\Bigl({d^2\over dr^2}+{2\over r}{d\over dr}\Bigr)C_m
+\sum_{i+j+k=m}C_iC_jC_k - 3C_mC^2 = -(J/D)\delta_{m1}\delta({\bf r}). \eqnoi
$$
{}From these equations one can subsequently find asymptotic solutions
for the total cluster concentration
$$
\Cr \simeq r^{-1}(4\logr)^{-1/2}, \eqnoi
$$
for the concentration of monomers
$$
C_1(r) \sim r^{-1}(\logr)^{-3/4}, \eqnoi
$$
etc. These results suggest the ansatz
$$
\Cmr = \Fmr/r, \eqnoi
$$
with $\rho = \logr$ for the behavior of $\Cmr$ for general $m$.

Substituting this ansatz into the governing equation, we obtain
$$
{dF_m\over d\rho} - {d^2F_m\over d\rho^2} =
\sum_{i+j+k=m}F_iF_jF_k - 3F_mF^2, \eqnoi
$$
with
$$
F=\sum\nolimits_{m=1}^{\infty}\Fmr=rC(r) \simeq (4\rho)^{-1/2}.
$$
Far away from the source, one can omit the second term
in the left-hand side of this equation.
After replacement of all $\Fmr$ by $\Cmt$, the
resulting approximate equation for source-induced aggregation
at the steady state becomes {\it identical} to Eq.~(12) for
irreversible aggregation, with $\rho$, $\rho = \logr$,
playing the role of time $t$. For the latter problem,
the cluster-mass distribution $\Cmt$ approaches the scaling
form not only in the monodisperse case but also for
arbitrary (rapidly vanishing) initial conditions [4]. In general
case, the scaling solution still has the form (15), where parameter
$N_0$ is equal to the total mass of the system,
$N_0=\sum\nolimits_{m=1}^{\infty}m\Cmt$ [4]. Since for the
former problem the density of mass $M(r)$,
$M(r)=\sum\nolimits_{m=1}^{\infty}m\Cmr$, satisfies
the Laplace's equation,
$$
D\Bigl({d^2\over dr^2}+{2\over r}{d\over dr}\Bigr)M(r)
= -J\delta({\bf r}), \eqnoi
$$
one gets $M(r)=J/4\pi Dr$. Therefore
$\sum\nolimits_{m=1}^{\infty}m\Fmr=rM(r)=J/4\pi D$ will play
the role of $N_0$ in the scaling solution. Combining all
these findings, we finally obtain that in the scaling limit
$$
m\to\infty, \quad r\to\infty, \quad
z=2\pi Dm/J\sqrt{\logr}=\rm finite, \eqnoi
$$
the steady-state cluster-mass distribution reaches the scaling form:
$$
C_{1+2m}(r)\simeq {\pi D\over Jr\logr}{\exp(-z)\over\ \sqrt{\pi z}}. \eqnoi
$$
Thus in the steady state in three dimensions, the typical mass
of clusters at distance $r$ from the source
grows unexpectedly slowly, $m_{\rm typ} \sim \sqrt{\logr}$.

{}From the steady-state (33), one can find the following limiting behavior
$$
C_{1+2m}(r)\simeq \sqrt{D\over J} m^{-1/2}r^{-1}(\logr)^{-3/4}, \eqnoi
$$
for $m \ll J\sqrt{\logr}/D$. Notice that in the steady state the total
cluster concentration $\Cr$ does not depend on the strength of the
source, $J$, while all other concentrations $\Cmr$ do depend on $J$.

For the general $n$-particle PCM, a complete asymptotic solution
for the steady-state cluster-mass distribution may be found at the
critical dimension $d=d^c$ following the procedure used for the
3-body case. By applying the ansatz $\Cmr = \Fmr/r^{d_c}$
one can recast the steady-state reaction-diffusion equation
to the the $n$-particle Smoluchowski equation. Making use of
the scaling solution of the latter equation (18), one obtains
$$
C_{1+(n-1)m}(r)\sim \Bigl[r\logr\Bigr]^{-d_c}z^{-{n-2\over n-1}}e^{-z}, \eqnoi
$$
with $z \sim m \logr^{-1/(n-1)}$ being the scaling variable.

\bigskip

In conclusion, we have obtained the complete asymptotic solution
for the steady-state cluster-mass distribution for the $n$-particle
aggregation model with a spatially localized source of monomers
in the critical dimension $d=d^c=2n/(n-1)$;
the binary case corresponds to $d=4$ while the ternary
case corresponds to $d=3$. The solution exhibits
an unexpectedly slow logarithmic growth of the typical
mass of clusters versus the distance $r$ from the source:
$m_{\rm typ}\sim \logr^{1/(n-1)}$.
For sufficiency large dimensions, $d>d^c$, we have found
that clusters do not interact far away from the source
and hence concentrations decay
diffusively as $r^{-(d-2)}$. Therefore for $d>d^c$ the
reaction leads to the renormalization of the strength
of the source but does not change the behavior qualitatively.
For sufficiently small dimensions, $d^c>d>d_c$,
the steady-state cluster-mass distribution is expected to be
of the scaling form $C_m(r) \sim r^{-\b}\Phi(mr^{-\a})$.
{}From two known moments of the cluster-mass distribution
$$
C(r)=\sum\nolimits_{m=1}^{\infty}\Cmr \sim r^{-d_c}
$$
[see Eq.~(21)] and
$$
M(r)=\sum\nolimits_{m=1}^{\infty}m\Cmr \sim r^{-(d-2)},
$$
one can find the exponents $\a$ and $\b$,
$\a=d^c-d$ and $\b=d_c+d^c-d$.
However the scaling function $\Phi(mr^{-\a})$ satisfies a
rather difficult integro-differential equation which
I could not solve. Further progress is also possible
in one dimension where exact solutions for the total
density of clusters and for the densities of clusters
of small mass may be readily found and behaviors even
in the vicinity of the source may be investigated.

\bigskip

I am grateful to D.~ben-Avraham for informing me on his results [7]
prior to publication and to E.~Ben-Naim, F.~Leyvraz, and S.~Redner
for useful discussions. I gratefully acknowledge ARO grant
\#DAAH04-93-G-0021 and NSF grant \#DMR-9219845 for partial
support of this research.

\vfill\eject
\centerline{\bf References}\smallskip

\refi for a review see, e.g., M.~H.~Ernst, in: {\it Fractals
in Physics}, L.~Pietronero and E.~Tosatti, eds. (Amsterdam, Elseveir,
1986) p.~289.
\refi K.~Kang and S.~Redner, \pra 32 435 1985 .
\refi K.~Kang, P.~Meakin, J.~H.~Oh, and S.~Redner, \jpa 17 L685 1984 .
\refi P.~L.~Krapivsky, \jpa 24 4697 1991 .
\refi S.~Cornell, M.~Droz, and B.~Chopard, \pA 188 322 1992 .
\refi V.~Privman and M.~D.~Grynberg, \jpa 25 6567 1992 .
\refi D.~ben-Avraham, \prl 71 3733 1993 .
\refi K.~Kang and S.~Redner, \pra 30 2833 1984 .
\refi J.~L.~Spouge, \prl 60 871 1988 .
\refi B.~R.~Thomson, \jpa 25 879 1989 .
\refi C.~R.~Doering and D.~ben-Avraham, \pra 38 3035 1988 ;
C.~R.~Doering and D.~ben-Avraham, \prl 62 2563 1989 ;
D.~ben-Avraham, M.~A.~Burschka and C.~R.~Doering \jsp 60 295 1990 .
\refi Z.~Cheng, S.~Redner, and F.~Leyvraz, \prl 62 2321 1989 .
\refi H.~Takayasu, \prl 63 2563 1990 ;
H.~Takayasu, M.~Takayasu, ~A.~Provata, and G.~Huber, \jsp 65 725 1991 .
\refi P.~L.~Krapivsky, \pA 198 150 1993 ;
P.~L.~Krapivsky, \pA 198 157 1993 .
\refi S.~Redner and D.~ben-Avraham, \jpa 23 L1169 1990 .
\refi P.~L.~Krapivsky, \pre 47 1199 1993 .
\refi M.~Bramson and D.~Griffeath, \zw 53 183 1980 .
\refi P.~Meakin and H.~Stanley, \jpa 17 L173 1984 ;
P.~Meakin, \pA 165 1 1990 .
\refi B.~P.~Lee, Renormalization group calculation for the
reaction $kA \to 0$, preprint.

\vfill\eject\bye

\end